\documentclass[pra,twocolumn,amssymb,superscriptaddress]{revtex4}
\usepackage{amsmath}
\usepackage{amsthm}
\usepackage{graphicx}
\usepackage[dvips]{epsfig}
\usepackage{bbm}
\usepackage{braket}
\usepackage[usenames,dvipsnames]{color}
\usepackage{wasysym}

\newcommand{\ie}{\emph{i.e.}}
\newcommand{\eg}{\textit{e.g.}}
\newcommand{\tr}{\operatorname{tr}}
\newcommand{\st}{\operatorname{s.t.}}
\newcommand{\prob}{\operatorname{Pr}}
\newcommand{\diag}{\operatorname{diag}}

\newcommand{\mean}[1]{\ensuremath{\langle #1 \rangle}}

\newtheorem{proposition}{Proposition}[section]
\newtheorem{example}{Example}[section]
\newtheorem{remark}{Remark}[section]

\newtheorem{definition}{Definition}[section]
\newtheorem{lemma}{Lemma}

\begin{document}
\title{Calibration robust entanglement detection beyond Bell inequalities}

\author{Tobias Moroder}
\affiliation{Institut f\"{u}r Quantenoptik und Quanteninformation, \"{O}sterreichische Akademie der Wissenschaften, Technikerstra{\ss}e~21A, A-6020~Innsbruck, Austria}
\author{Oleg Gittsovich}
\affiliation{Department of Physics \& Astronomy, Institute for Quantum Computing University of Waterloo, 200~University Avenue West, N2L~3G1~Waterloo, Ontario, Canada}

\date{\today}
\begin{abstract}
In its vast majority entanglement verification is examined either in the complete characterized or totally device independent scenario. The assumptions imposed by these extreme cases are often either too weak or strong for real experiments.  Here we investigate this detection task for the intermediate regime where partial knowledge of the measured observables is known, considering cases like orthogonal, sharp or only dimension bounded measurements. We show that for all these assumptions it is not necessary to violate a corresponding Bell inequality in order to detect entanglement. We derive strong detection criteria that can be directly evaluated for experimental data and which are robust against large classes of calibration errors. The conditions are even capable of detecting bound entanglement under the sole assumption of dimension bounded measurements.
\end{abstract}

%\pacs{03.67.Mn, 03.65.Ud, 03.65.Ta}
\maketitle

\section{Introduction}
Entanglement is the most striking phenomenon in the quantum world. It provides the resource for fascinating new applications such as quantum computing, teleportation or unconditional secure communiction. These possibilities drastically sparked the interest for this resource and many current experiments strive to realize strong and robust entanglement. Consequently many different detection methods have been developed in recent years, for reviews see Refs.~\cite{horodecki_review,guehne09a}.

Reliable entanglement verification must fulfil certain criteria~\cite{vanenk07a}. Most importantly it should not depend on the preparation procedure and the only available information about the presence of entanglement should be obtained via measurements of the underlying system. However there is still one open choice left: Usually each classical outcome is associated with an operator describing the measurement apparatus, say outcome ``k'' corresponds to the measurement operator $M_k$. Equipped with this quantum mechanical meaning one can employ for instance the tool of entanglement witnesses~\cite{horodecki96b,terhal00a,doherty04a} to decide the entanglement question. This standard scenario might be too optimistic for applications because it crucially relies on the correctness of the employed operator description of the measuring device and clearly if the true measurement apparatus functions are different then anything can go wrong. For mere entanglement detection these deviations might be called systematic errors but for applications where the presence of entanglement is essential, secure communication being prominent example, these deviations are undesired pitfalls~\cite{qi07a,lydersen10a}. Thus in contrast to the completely characterised scenario there is also the other extreme where one does not need any specific quantum mechanical model at all. Though surprising at first even in this totally pessimistic device independent case it is possible to infer entanglement for good enough data, for example using Bell inequalities~\cite{bell64a,clauser74a,peres99a}.

Detection of entanglement in a complete device independent manner has recently attracted a lot of interest; in particular verification in the multipartite settings~\cite{bancal11a,pal11a,brunner11a} since it was realized that this task differs from the corresponding non-locality one~\cite{bancal11a}. This was surprising because device independent entanglement detection and exclusion of a local hidden variable model are equivalent problems in the bipartite case~\cite{werner89a,acin06a}. Steering inequalities are entanglement detection methods in a hybrid scenarios, \ie, one party is complete characterised, the other totally uncharacterised~\cite{wiseman07a}. Only a few results and techniques have addressed the detection of entanglement in a partially characterized setting so far. References~\cite{roy05a,uffink08a} considered the case of sharp, orthogonal qubit measurements and showed that much more entangled state can in fact be detected than with the corresponding Bell inequality. For instance it was proven that the bound appearing in the famous Bell correlation term of the CHSH inequality~\cite{clauser74a} can be actually reduced from $2$ to $\sqrt{2}$ when the measurements satisfy this extra constraint. These results have been extended in order to provide even quantitative bounds on the amount of entanglement in Ref.~\cite{lougovski09a}. In order to devise more robust entanglement detection methods that avoid fake entanglement detection under misspecification of the employed observables a technique called squash model is very useful \cite{beaudry08a,tsurumaru08a}. Applied to entanglement verification, this notion, usually common in quantum key distribution, can even be enlarged~\cite{moroder10a}. Finally, Ref.~\cite{shchukin09a} introduces types of Bell inequalities that can be applied if the commutator of different measurement settings is known.

The purpose of the current manuscript is to investigate the verification task in the intermediate regime where one possesses some partial knowledge of the employed measurement devices. Despite its elegance the complete device independent setting suffers from the drawback that it is hard to address experimentally. In fact all Bell experiments so far were not fully device independent due to certain loopholes, but which are often irrelevant if certain other assumptions hold. However such assumptions like the fair sampling condition effectively can be regarded as partial information, \eg, that the ``inconclusive'' measurement outcome is the same for both settings for the fair sampling case. But clearly if these additional assumptions are not satisfied, the corresponding implementation of the Bell test also does not provide a positive entanglement check~\cite{semenov08a,gerhardt11a}. Besides, the complete device independent scenario is often also too pessimistic. Of course for applications like quantum key distribution this very pessimistic viewpoint is legitimate but for the mere verification of entanglement in an experiment this seems like breaking a butterfly on a wheel. Although we do not address the question how such partial knowledge can be obtained or justified, we nevertheless believe that certain deviations in a measurement description are more stringent than others. For example the assumption that the measurement of the electronic state of an ion in a trap is very well described by a qubit measurement seems much easier to assure than the assumption that the performed measurements are really orthogonal or that they are true projectors. Independent of this discussion which scenario is now more reasonable for which situation, our investigation shed some light on the question which assumptions are more crucial than others in order to verify entanglement. In addition, the derived entanglement criteria are more robust against calibration errors while they still keep a large detection strength, in particular when compared to Bell inequalities.

In the following we first focus on the scenario investigated in Ref.~\cite{uffink08a} and analyse entanglement verification in the simplest possible setting of two different dichotomic measurement settings per side. We draw our attention to different assumptions about qubit measurements and distinguish three different classes: sharp, orthogonal or the completely uncharacterised qubit measurements. We provide a solution in terms of the singular values of a corresponding data matrix and find that one detects a much larger fraction of states than with the complete device independent setting. This already provides examples where one detects entanglement although the corresponding Bell inequalities, the CHSH inequalities \cite{clauser74a} in this case, are not violated with the sole extra assumption that the dimension of the underlying quantum system is fixed. Additionally we consider specific observations where the additional knowledge of sharpness or orthogonality is irrelevant for the detection strength and already the dimension restriction suffices to verify exactly the same amount of entangled states as with completely characterized, \ie, sharp and orthogonal measurements. After considering these various scenarios for two qubits we focus on the extensions to more dichotomic measurements with completely unspecified measurements restricted only by the underlying dimension. We derive a criterion that is applicable for any of these settings and show that it is capable of detecting entanglement in data originating from bound entangled states. Moreover the criterion even shows that with uncharacterised qutrit measurements one can verify more entanglement than with corresponding Bell inequalities. Note that dimensions $d\geq 4$ are not relevant, because then one effectively equals the detection strength of the Bell inequalities \cite{werner89a,acin06a}.

The outline is as follows: In Sec.~\ref{sec:II} we precisely define the entanglement verification under partial information on the performed measurements. Section~\ref{sec:III} starts with a discussion about different assumptions on the measurements. In addition we provide some further notation and background knowledge about the entanglement criterion that we employ for our purpose. Section~\ref{sec:IIImain} finally contains the above mentioned results for the two-qubit case, whereas Sec.~\ref{sec:IV} is devoted to the general scenario of $n$ uncharacterized dichotomic qudit measurements. Finally we conclude and comment on possible further extensions and directions in Sec.~\ref{sec:V}. Some technical details of the proofs can be found in appendix.

\section{Problem definition}\label{sec:II}

Suppose that Alice and Bob observe an outcome probability distribution denoted as $\prob(x,y|a,b)$, where $a$ labels different measurement choices with respective outcomes~$x$ for Alice and similar for Bob. These observed data have a quantum mechanical representation if there exists a quantum state $\rho_{\rm AB}$ and corresponding measurements, \ie, set of positive operator-valued measures (POVM) for Alice $\{ M_x^a\}$ and $\{M_y^b \}$ for Bob such that
\begin{equation}
\prob(x,y|a,b)=\tr(\rho_{\rm AB} M_x^a \otimes M_y^b), \;\forall x,y,a,b.
\end{equation}
The observed data are said to verify entanglement if and only if all states $\rho_{\rm AB}$ that satisfy this relation are entangled. Note that the measurement description is very crucial here because it ties a quantum mechanical meaning to the classical outcomes.

In the following we consider the alternative that only partial knowledge is possessed on the measurement description, meaning that the POVMs describing the measurement are not known completely. Hence each local measurement characterization is only assured to lie within a certain class. This set of possible POVMs will be denoted by $\mathcal{M}_{\rm A}$ for Alice and $\mathcal{M}_{\rm B}$ for Bob. In this case, successful entanglement detection implies that for all measurement descriptions only entangled states give rise to the observed data. More precisely, if $\mathcal{S}$ denotes the set of states having a quantum representation in accordance with the assumed measurement description,
\begin{align}
\label{eq:defS}
\mathcal{S}= \{ \rho_{\rm AB}  &|   \exists \{M_x^a\} \in \mathcal{M}_{\rm A},  \{M_y^b\} \in \mathcal{M}_{\rm B}: \\
\nonumber
& \prob(x,y|a,b)=\tr(\rho_{\rm AB} M_x^a \otimes M_y^b), \;\forall x,y,a,b \},
\end{align}
then the observed data $\prob(x,y|a,b)$ certify entanglement if and only if all state of this set $\mathcal{S}$ are entangled.

Let us comment on the two extreme cases: If the measurements are completely specified each set $\mathcal{M}$ only consists of one possible element. In such case the question whether given observations verify entanglement is completely answered for example with the help of entanglement witnesses, even if the measurement does not provide full tomography~\cite{curty04a}. In the other extreme that the measurements are completely unspecified the sets $\mathcal{M}$ consist of all possible POVMs in all possible dimensions and the data exclusively correspond to entangled states if and only if a Bell inequality with the specified number of settings and outcomes is violated~\cite{werner89a}.

Finally let us stress one more technical point: We do not assume any ``convexification'' of the problem as for example employed in Ref.~\cite{brunner08a} for dimension witnesses. Convexification would mean that if two different observations $P_1, P_2$ would have a separable quantum representation then also its convex combination $\lambda P_1 + (1-\lambda) P_2$ for all $\lambda \in [0,1]$. However the problem is that the quantum representations might need different measurements, so that directly taking the convex combination on the level of quantum states does not work anymore. In Sec.~\ref{sec:IIImain} we provide an explicit counterexample, where convexification would lead to spurious entanglement detection.

\section{Qubit case}\label{sec:III}

This section concentrates on the two-qubit scenario. We start with the definition of different measurement assumptions followed by an explicit parametrization. Afterwards we introduce the notation of a data matrix in order to express our results more compactly and also state the entanglement criterion that is employed to prove the main results in the last part of this section.

\subsection{Different measurement assumptions}

At first, let us specify the different measurement properties more closely which were abstractly described by the set $\mathcal{M}$ in the previous section. We consider the simplest non-trivial case: Each party has two different measurement settings each of which having two different outcomes. Any of such dichotomic measurements is more compactly determined by the difference of two POVM elements, \eg, for the first setting of Alice we associate the operator
\begin{equation}
\label{eq:diff2povm}
A=M^a_{+1} - M^a_{-1}  \;\; \Leftrightarrow \;\; M^a_{\pm 1} =\frac{1}{2} \left( \mathbbm{1} \pm A \right).
\end{equation}
Here $x=\pm 1$ labels the two different outcomes, while the resolution of the POVM elements $M_{\pm 1}^a$ follows because of normalization. In order that this operator $A$ describes a valid POVM it must satisfy the conditions,
\begin{equation}
\label{eq:condpovm}
A - \mathbbm{1}\geq 0, \; \mathbbm{1}-A\geq 0.
\end{equation}
Let us remark that this condition is still independent of any dimension restriction and it will reappear in the later section for the more general case of uncharacterised qudit measurements. The operator for the second choice of Alice is denoted by $A^\prime$, while Bob's choices are given by $B,B^\prime$ respectively.

The following definition summarizes the different qubit specifications that we consider. Let us point out that the dimension restriction seems to us as the first non-trivial assumption that one can make \footnote{The only other alternative would be to provide a distance measure for the set of POVMs, \ie, $\delta( \{ M_{x,\rm ideal}^a \}, \{M_{x,\rm true}^a \}) \leq \epsilon$ which quantifies the difference between the true and the ideal measurement description. If one wants to make this bound independent of the dimension this norm must be independent of the dimension as well. However we cannot think of any reasonable distance here.}.

\begin{definition}[Qubit measurement models]\label{def:models}
For two dichotomic measurements, characterized by the operators $A,A^\prime$ according to Eq.~(\ref{eq:diff2povm}) and satisfying Eq.~(\ref{eq:condpovm}), we distinguish the following cases:
\begin{enumerate}
\item[i)] Qubit measurements: Both operators act on \emph{the same} qubit.
\item[ii)] Sharp qubit measurements: The POVM elements are rank-1 projectors on the same qubit, \ie, $A, A^\prime$ have eigenvalues $\pm 1$.
\item[iii)] Orthogonal qubit measurements: The eigenbasis of $A$ and $A^\prime$ are mutually unbiased \footnote{Note that this does not imply the orthogonality of $A$ and $A'$ with respect to Hilbert-Schmidt norm.}.
\end{enumerate}
\end{definition}

\subsection{Parametrization of POVM elements}\label{sec:parametrization}

In the following we introduce a parametrization of the POVM elements corresponding to different measurement scenarios. This parametrization will be convenient later for the technical proofs. Additionally it should further clarify the different measurement properties.

\subsubsection{Sharp but not orthogonal qubit measurements}
In this case the operators $A,A^\prime$ can be written as follows:
\begin{eqnarray}
A &=& \cos(\theta) \sigma_i + \sin(\theta) \sigma_j \\
A^\prime &=& \cos(\theta) \sigma_i - \sin(\theta) \sigma_j,
\end{eqnarray}
where $\sigma_i, \sigma_j$ are two different possibly rotated Pauli operators, \ie, they can be written as $\sigma_i=\hat u_i \cdot \vec \sigma$ and $\sigma_j=\hat u_j \cdot \vec \sigma$ with two unit vectors $\hat u_i, \hat u_j \in \mathbbm{R}^3$ satisfying $\hat u_i \cdot \hat u_j = 0$. The parameter $\theta$ characterizes the tilt between the measurement directions. Note that this relation can also be reversed, \ie, to express the Pauli operator in terms of the considered measurement operators. Formally the relation between orthogonal and non-orthogonal observables is described by
\begin{equation}
\left[ \begin{array}{ccc} \mathbbm{1} \\ \sigma_i \\ \sigma_j \end{array} \right]= R(\theta) \left[ \begin{array}{ccc} \mathbbm{1} \\ A \\ A^\prime \end{array} \right]
\end{equation}
with~\footnote{For notation clarity if matrix entries are not visualized they are equal to zero, if they can be arbitrary it is symbolized by $\ast$.}
\begin{equation}
\label{eq:R3}
R(\theta)= \left[ \begin{array}{ccc} 1 & & \\ & \frac{1}{2\cos(\theta)} & \frac{1}{2\sin(\theta)} \\ & \frac{1}{2\cos(\theta)} & -\frac{1}{2\sin(\theta)} \end{array} \right].
\end{equation}
If we only refer to the $2 \times 2$ sub-matrix, formed by the second and third column and row, we use the label $R_2(\theta)$.

\subsubsection{Non-sharp but orthogonal qubit measurements}
When the measurements are not sharp but orthogonal, we directly employ the reverse parametrization
\begin{eqnarray}
\label{eq:unsharpMeas1}
\sigma_i &=& x_1 \mathbbm{1} + x_2 A \\
\label{eq:unsharpMeas2}
\sigma_j &=& x_3 \mathbbm{1} + x_4 A^\prime,
\end{eqnarray}
with $x_i \in \mathbbm{R}$. Note that in order that $A,A^\prime$ correspond to physical observables given by Eq.~(\ref{eq:condpovm}) these parameters must satisfy $x_2 \geq 1 + |x_1|$ and $x_4 \geq 1 + |x_3|$. Here we can choose without loss of generality $x_2,x_4 > 0$ to be positive, by selecting appropriately $\sigma_i$ or $-\sigma_i$. Formally sharp and non-sharp observables can be related by
\begin{equation}
\left[ \begin{array}{ccc} \mathbbm{1} \\ \sigma_i \\ \sigma_j \end{array} \right]= S(\vec x) \left[ \begin{array}{ccc} \mathbbm{1} \\ A \\ A^\prime \end{array} \right]
\end{equation}
with
\begin{equation}
\label{eq:Sofx}
S(\vec x)= \left[ \begin{array}{ccc} 1 & &\\ x_1 & x_2 & \\ x_3 & & x_4 \\\end{array} \right].
\end{equation}

\subsubsection{Uncharacterised qubit measurement}
The remaining case of totally uncharacterised qubit measurements can be considered as a combination of the above two cases. The overall transformation is given by
\begin{equation}
\left[ \begin{array}{ccc} \mathbbm{1} \\ \sigma_i \\ \sigma_j \end{array} \right]= R(\theta) S(\vec x) \left[ \begin{array}{ccc} \mathbbm{1} \\ A \\ A^\prime \end{array} \right].
\end{equation}
The first operation $S$ turns the operators $A,A^\prime$ into sharp, but not necessarily orthogonal measurements, which is considered afterwards by applying the transformation $R$.

\subsection{Data matrix}

In order to express our results let us define some further notation. The observed data $\prob(x,y|a,b)$ are compactly expressed in terms of a data matrix $D_3$, given by the matrix of expectation values
\begin{equation}
D_3=\left[\begin{array}{ccc}
\mean{\mathbbm{1}} & \mean{B} & \mean{B^\prime} \\
\mean{A} & \mean{AB} & \mean{AB^\prime} \\
\mean{A^\prime} & \mean{A^\prime B} & \mean{A^\prime B^\prime} \end{array} \right].
\end{equation}
For convenience we often abbreviate the $2\times 2$ sub-matrix containing only the full correlations, \ie, built up by the second and third row and column, as $D_2$. Our criteria are typically given in terms of the singular values of this sub-matrix, denoted as $\lambda_{1/2} \geq 0$.

\subsection{Employed entanglement criteria}

For entanglement detection we employ a criterion, which is a direct corollary of the computable cross-norm or realignment (CCNR) criterion~\cite{rudolph02a,chen03a}. The corollary is formulated in terms of the singular values of the correlation matrix $T_3$ given by
\begin{equation}
T_3=\left[\begin{array}{ccc}
\mean{\mathbbm{1}} & \mean{\sigma^{\rm B}_i} & \mean{\sigma^{\rm B}_j}\\
\mean{\sigma^{\rm A}_i} & \mean{\sigma^{\rm A}_i \sigma^{\rm B}_i} & \mean{\sigma^{\rm A}_i \sigma^{\rm B}_j} \\
\mean{\sigma^{\rm A}_j} & \mean{\sigma^{\rm A}_j \sigma^{\rm B}_i} & \mean{\sigma^{\rm A}_j \sigma^{\rm B}_j} \end{array} \right].
\end{equation}
Note that this correlation matrix $T_3$ represents a special data matrix $D_3$ for which the employed measurement operators are sharp and orthogonal. Because of those similarities we employ the similar label $T_2$ in order to refer to the full correlation $2\times 2$ sub-matrix.

\begin{proposition}[Corollary of the CCNR criterion]\label{prop:ent_criterion}
Given the correlation matrix $T_3$ with ordered singular values $\lambda_0 \geq \lambda_1 \geq \lambda_2 \geq 0$. Then the CCNR criterion implies that any separable state necessarily satisfies
\begin{equation}
\label{eq:ent_criterion}
\|T_3\|_1 = \lambda_0 + \lambda_1 + \lambda_2 \leq 2.
\end{equation}
If the correlation matrix has vanishing marginals, \ie, $\mean{\sigma^{\rm A}_k}=\mean{\sigma^{\rm B}_k}=0$ for all $k \in \{i,j\}$, and $\lambda_0=1$ then this condition is also sufficient.
\end{proposition}

For completeness we provide a proof of this proposition in App.~\ref{proof:ent_criterion}. With this stage set we will state in the next section our main results on entanglement verification in different two qubit scenarios.

\subsection{Main results}\label{sec:IIImain}

\begin{figure}[t]
\centering
\includegraphics[scale=.95]{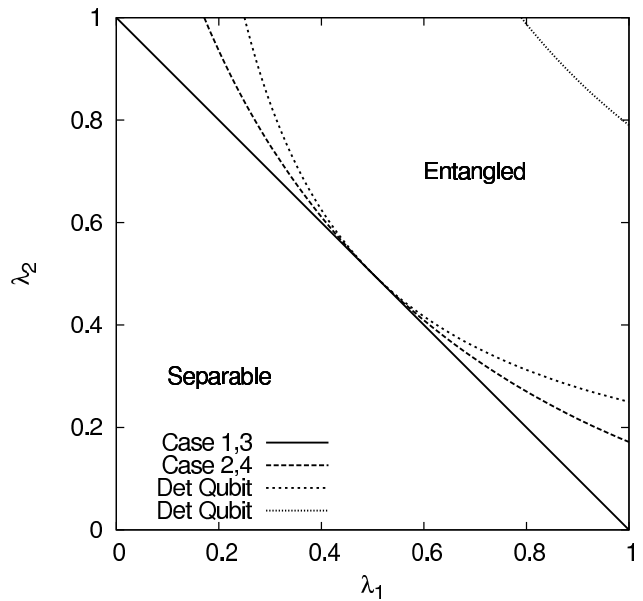}
\caption{Different detection regions for a data matrix with singular values $\lambda_1,\lambda_2$. The solid and long-dashed lines correspond to the case of a data matrix with vanishing marginals as discussed in Prop.~\ref{prop:main_result}. The remaining two lines correspond to determinant detection rule $\det(D)=\lambda_1 \lambda_2$ given in Prop.~\ref{prop:detectMe} for qubits (dotted) and qutrits (short-dashed).}
\label{fig:detectioin_regions}
\end{figure}

In the following we state and prove our main results for qubits. We first consider the special case that the observed data matrix $D_3$ has vanishing marginals. We obtain a complete solution for different scenarios if one only uses the knowledge of the appearing singular values, \ie, the criteria are minimized over all data matrices with fixed singular values. Using additional structure of the observation improves the detection strength as we will see later in Prop.~\ref{prop:D3diag}. For comparison reason the different detection regions are visualized in Fig.~\ref{fig:detectioin_regions}.

\begin{proposition}(Data matrix with zero marginals)\label{prop:main_result}
Given a data matrix $D_3$ of the following form
\begin{equation}
D_3 = \left[ \begin{array}{cc} 1& \\ & D_2 \end{array} \right],
\end{equation}
where the full correlation data matrix $D_2$ is characterized by the singular values $\lambda_1 \geq \lambda_2 \geq 0$. These data verify entanglement under the assumption that both measurement of Alice and Bob are
\begin{enumerate}
\item[1)] Sharp and orthogonal: $\lambda_1 + \lambda_2 >1$
\item[2)] Sharp, non-orthogonal: $\sqrt{\lambda_1} + \sqrt{\lambda_2} > \sqrt{2}$
\item[3)] Unsharp, orthogonal: $\lambda_1 + \lambda_2 >1$
\item[4)] Qubit measurements: $\sqrt{\lambda_1} + \sqrt{\lambda_2} > \sqrt{2}$
\end{enumerate}
The definition of these properties is given in Def.~\ref{def:models} and these bounds are tight for the considered scenario. 
\end{proposition}
\begin{remark}
Note that we assume that $D_3$ actually originates from a quantum state under the considered measurement scenario, which can be assured for example if the singular values satisfy $1\geq \lambda_1\geq \lambda_2 \geq 0$.
\end{remark}

\begin{proof}
Case 1) of sharp and orthogonal measurements has already been discussed in Ref.~\cite{uffink08a}. Alternatively it is a direct application of Prop.~\ref{prop:ent_criterion}. 

All other scenarios are proven along the following lines: Given the data matrix $D_3$ one first reconstructs the corresponding correlation matrix $T_3$ by the appropriate transformations $S(\vec x),R(\theta)$ as given in Sec.~\ref{sec:parametrization}. In order to certify entanglement one employs Prop.~\ref{prop:ent_criterion}. However, since $T_3$ depends on the transformation parameters, \eg, $\theta, \vec x,\dots$, one needs to optimize over all such choices. This will in general result in lower bounds on the singular values of the data matrix. If these bounds are tight then the provided condition is necessary and sufficient in order to detect entanglement with provided data.

Case 2) At first let us concentrate on the sharp but non-orthogonal case, in which the correlation matrix is given by $T_3=R(\alpha)D_3 R(\beta)^T$. For the block-diagonal data matrix the resulting correlation matrix is of similar block structure, \ie, $T_3=\diag[1,T_2]$ with
\begin{equation}
\label{eq:t2direct}
T_2 = R_2(\alpha) D_2 R^T_2(\beta).
\end{equation}
Hence if the ordered singular values of $T_2$ are denoted as $t_1 \geq t_2\geq 0$ then Prop~\ref{prop:ent_criterion} states that the state is entangled if and only if $t_1 + t_2 > 1$ holds for all values of $\alpha,\beta$. In order to minimize the sum $t_1 + t_2$ over the angles we first lower bound this quantity by an expression containing only the singular values of the appearing transformations since this is more easily optimized in the end.

The lower bound is derived using the inverse relation of Eq.~(\ref{eq:t2direct}),
\begin{equation}
\label{eq:D2-1}
D_2 = R_2(\alpha)^{-1} T_2 R^T_2(\beta)^{-1},
\end{equation}
with
\begin{equation}
\label{eq:R2-1}
R_2(\alpha)^{-1}= \left[ \begin{array}{cc} \cos(\alpha) & \sin(\alpha) \\ \cos(\alpha) & -\sin(\alpha) \end{array} \right].
\end{equation}
In the following discussion we employ the abbreviations $a_1\geq a_2\geq 0$ and $b_1\geq b_2\geq 0$ for the ordered singular values of $R^T_2(\alpha)^{-1}$ and $R^T_2(\beta)^{-1}$ respectively. Furthermore recall that $D_2$ is characterized by its two singular values $\lambda_1\geq \lambda_2\geq 0$. For the matrices on the left- and right-hand side of Eq.~(\ref{eq:D2-1}) the following relations hold
\begin{eqnarray}
\label{eq:help1}
t_1 t_2 &=& \frac{\lambda_1 \lambda_2}{a_1 a_2 b_1 b_2}, \\
\label{eq:help2}
t_1^2+t_2^2 &\geq& \frac{(\lambda_1 + \lambda_2)^2}{a_1^2b_1^2 + a_2^2b_2^2}.
\end{eqnarray}
The proof of these two relations involves some technical details and is given in App.~\ref{sec:proof_help12}. Employing these two identities provide
\begin{align}
(t_1+t_2)^2 &\geq \min_{\alpha,\beta} \left[ \frac{(\lambda_1 + \lambda_2)^2}{a_1^2b_1^2 + a_2^2b_2^2} +  2 \frac{\lambda_1 \lambda_2}{a_1 a_2 b_1 b_2}\right] \\
\label{eq:solutionSNO}
& \geq  \frac{1}{4} \left(\sqrt{\lambda_1} + \sqrt{\lambda_2} \right)^4.
\end{align}
The last inequality arises if one employs the true singular values $a_1(\alpha),\dots$ and performs the minimization; for an explicit proof of this optimization see Lemma~\ref{lemma1} in App.~\ref{sec:app}. Eq.~(\ref{eq:solutionSNO}) confirms that the state is entangled if and only if $\sqrt{\lambda_1} + \sqrt{\lambda_1} > \sqrt{2}$, where the sufficiency follows from the fact that all appearing inequalities can also be achieved with equality. This finilizes the proof of the Case 2).

Case 3) Next consider the unsharp, orthogonal case. The correlation matrix $T_3=S(\vec x) D_3 S(\vec y)^T$, with appropriate ``sharpener'' transformations $S(\vec x)$ given by Eq.~(\ref{eq:Sofx}), has the following block structure
\begin{align}
T_3 &= \left[ \begin{array}{cc} 1 & \\  x & S_x \end{array}\right] \left[ \begin{array}{cc} 1 & \\  & D_2 \end{array}\right] \left[ \begin{array}{cc} 1 & y^T \\  & S_y^T \end{array}\right] \\
&= \left[ \begin{array}{cc} 1 & y^T \\ x & xy^T + S_x D_2 S_y^T \end{array}\right].
\end{align}
Here we used an analogue block decomposition for $S(\vec x)$ with column vector $x=[x_1,x_3]^T$ and diagonal sub-matrix $S_x=\diag[x_2,x_4]$ and respective abbreviations for the transformation of Bob. In order to prove the statement we will use of the following three inequalities for the ordered singular values of $T_3$,
\begin{align}
\label{eq:cond1}
t_0  &\geq 1, \\
\label{eq:cond2}
t_0 t_1  &\geq \lambda_1, \\
\label{eq:cond3}
t_0 t_1 t_2  &\geq \lambda_1 \lambda_2.
\end{align}
The proof of this inequalities is given in App.~\ref{sec:proof_conds123}. Further, as shown in Lemma~\ref{lemma2} of App.~\ref{sec:app} these conditions, together with the ordering condition $1 \geq \lambda_1 \geq \lambda_2\geq 0$, ensure
\begin{equation}
t_0 + t_1 + t_2 \geq 1 +\lambda_1 +\lambda_2.
\end{equation}
If this ordering is not valid, \ie, $\lambda_1>1$, entanglement directly follows because of
\begin{equation}
t_0 + t_1 + t_2 \geq t_0 +t_1 \geq 2 \sqrt{t_0 t_1} >2
\end{equation}
via the inequality of arithmetic and geometric means. Thus in total $\lambda_1 +\lambda_2 > 1$ is necessary and sufficient for entanglement, sufficiency because one detects the same as in the more restrictive case of sharp, orthogonal measurements. This finishes the proof for the non-sharp, orthogonal case of qubit measurements.

Case 4) For the remaining scenario of fully uncharacterised qubit measurements we can largely employ the previous results. In this scenario the correlation matrix $T_3=R(\alpha) S(\vec x) D_3 S(\vec y)^T R(\beta)^T$ is given by
\begin{align}
\!  T_3 &= \left[\! \begin{array}{cc} 1 & \\ & R_2(\alpha) \end{array} \! \right] \! \! \left[  \! \begin{array}{cc} 1 & y^T \\ x & xy^T + S_x D_2 S_y^T \end{array}  \! \right] \! \! \left[ \! \begin{array}{cc} 1 & \\ & R_2(\beta)^T \end{array}  \! \right]\\
\label{eq:general_T3}
&= \left[ \begin{array}{cc} 1 & \tilde y^T \\ \tilde x & \tilde x \tilde y^T + R_2(\alpha) S_x D_2 S_y^T R_2(\beta)^T \end{array}\right]
\end{align}
with $\tilde x = R_2(\alpha) x, \tilde y = R_2(\beta)y$. In this case the important sub-matrix is
\begin{equation}
\label{eq:general_T2bar}
\bar T_2= R_2(\alpha) S_x D_2 S_y^T R_2(\beta)^T = R_2(\alpha) \bar D_2 R_2(\beta)^T,
\end{equation}
which can be considered as the central sub-matrix of the sharp but non-orthogonal case, Eq.~(\ref{eq:t2direct}), but where the transformation is applied to $\bar D_2$ instead of the true data matrix $D_2$ itself. From the sharp, non-orthogonal case we know that the singular values of this matrix $\bar T_2$, denoted as $\bar t_1 \geq \bar t_2$ satisfy
\begin{equation}
\label{eq:inequ1}
\bar t_1 + \bar t_2 \geq \frac{1}{2} \left( \sqrt{\bar \lambda_1} + \sqrt{\bar \lambda_2 }\right)^2 \geq \frac{1}{2} \left( \sqrt{\lambda_1} + \sqrt{\lambda_2 }\right)^2,
\end{equation}
where $\bar \lambda_i$ are singular values of $\bar D_2$. Next, using similar arguments as already presented in the sharp but non-orthogonal case we can derive the following set of inequalities for the singular values of $T_3$:
\begin{align}
t_0  &\geq 1, \\
t_0 t_1  &\geq  \bar t_1, \\
t_0 t_1 t_2  & \geq \bar t_1 \bar t_2.
\end{align}
Using once more Lemma~\ref{lemma2} and Eq.~(\ref{eq:inequ1}) provides
\begin{equation}
t_0 + t_1 +t_2 \geq 1 + \bar t_1 + \bar t_2 \geq 1 + \frac{1}{2}\left( \sqrt{\lambda_1} + \sqrt{\lambda_2 } \right)^2
\end{equation}
if one has the ordering $1\geq \bar t_1 \geq \bar t_2 \geq 0$. If $\bar t_1 > 1$ one verifies entanglement again by the inequality of arithmetic and geometric means. This finally shows that the state is entangled if and only if $\sqrt{\lambda_1} + \sqrt{\lambda_2 } > \sqrt{2}$ which proves the claim for uncharacterised
qubit measurements.
\end{proof}
 
Next let us provide an important numerical example. First it demonstrates that the unsharp, orthogonal case is indeed different from the completely characterized case. Together with the previous proposition it shows that the sharp, non-orthogonal and unsharp, orthogonal case are indeed inequivalent to each other, \ie, there are observations which are exclusively detected by one of these two scenarios. Additionally, this example proves that the entanglement verification in the unsharp case is in fact a non-convex problem. This means that one must be very careful in applying Prop.~\ref{prop:main_result}, it is for example not possible to use it on a ``depolarized'' version of the observed data matrix $D_3$, \ie, the one that one obtains by setting the marginals equal to zero.

\begin{example}
The data matrix
\begin{align}
\label{eq:exD3}
D_3 &= \left[ \begin{array}{ccc} 1& 1-\sqrt{3} & \\ 1-\sqrt{3} & (15-8\sqrt{3})/2 & \\ & & 1/2 \end{array} \right] \\
   &\approx \left[ \begin{array}{ccc} 1& -0.73 & \\ -0.73 & 0.57 & \\ & & 0.5 \end{array} \right]
\end{align}
can originate from a separable state in the case of unsharp orthogonal measurements, but verifies entanglement for sharp, non-orthogonal measurements. 

Moreover it shows that the unsharp scenarios are indeed non-convex problems, \ie, convex combination of two separable data matrices might not be separable anymore.
\end{example}

\begin{proof}
First let us give the separable state and its corresponding measurements that are consistent with the given data matrix. This also discloses the generation of this example. The data matrix given by Eq.~(\ref{eq:exD3}) is obtained by measuring the separable state
\begin{equation}
\rho_{\rm sep} =\frac{1}{4} \left[ \mathbbm{1} \otimes \mathbbm{1} + \frac{1}{2} \left( \sigma_x\otimes \sigma_x + \sigma_z\otimes \sigma_z \right)\right]
\end{equation}
with unsharp measurements, $A$ parametrized according to Eq.~(\ref{eq:unsharpMeas1}) with $\sigma_i=\sigma_x$, $x_1=1+\sqrt{3}$ and $x_2=1+x_1$, while $A^\prime = \sigma_z$ and the same settings for Bob's side. The reason for the choice of $x_1$ becomes clear afterwards.

Verifying entanglement under the premise of sharp qubit observables goes as follows: Note that if the measurements are even orthogonal applying the CCNR criterion of Prop.~\ref{prop:ent_criterion} immediately shows entanglement. In the non-orthogonal case one utilizes again the transformations $R(\alpha), R(\beta)$ in order to generate $T_3$ as done in the previous proposition. Even with non-vanishing marginals the important sub-matrix is given by $T_2=R_2(\alpha)D_2 R_2(\beta)^T$ with $D_2$ being the sub-matrix containing the full correlations. Though not directly stated as the CCNR criterion, a state is already entangled if the singular values of $T_2$ satisfy $t_1+t_2 >1$ \footnote{If the correlation matrix $T_3$ corresponds to a separable state, then also $\bar T_3$ where the marginals have been inverted. Since correlation matrices of separable states form a convex structure this assures that also the depolarized version $\tilde T_3=(T_3+\bar T_3)/2=\diag[1,T_2]$ is separable. Applying the CCNR criterion to $\tilde T_3$ assures entanglement if $t_1+t_2>1$ is fulfilled.}. As shown in the proof of the previous proposition this relation is assured if the singular values of $D_2$ fulfil $\sqrt{\lambda_1} + \sqrt{\lambda_2} > \sqrt{2}$. Since the data matrix given by Eq.~(\ref{eq:exD3}) satisfies this condition this proves that $D_3$ cannot be compatible with a separable state under sharp qubit measurements.

Finally one needs to verify that the data matrix given by Eq.~(\ref{eq:exD3}) is at all consistent with a valid quantum state using sharp measurements. This is necessary in order to assure that the set $\mathcal{S}$ defined as in Eq.~(\ref{eq:defS}) is indeed non-empty. However the operator
\begin{equation}
\rho_{\rm ent} =\frac{1}{4} \Big( y \sigma_y \otimes \sigma_y + \sum_{i,j\in \{0,x,y\}} [D_3]_{ij} \sigma_i\otimes \sigma_j \Big)
\end{equation}
with $y=4\sqrt{3}-7$ represents a valid state compatible with the data matrix if one employs the sharp measurements $A=B=\sigma_x,A^\prime=B^\prime =\sigma_z$. Let us point out that this physicality condition determined the parameter $x_1$: We optimized the detection condition $\sqrt{\lambda_1} + \sqrt{\lambda_2}$ while still keeping the data compatible with a valid state.

Furthermore, this example provides an explicit instance for the failure of convexity~\footnote{Note that this is not shown via Prop.~\ref{prop:main_result} because one only employ some restricted information from the observations, solely knowledge of the singular values.}. If $D_3$ corresponds to a separable state then also its marginal inverted version $\bar D_3$ because it effectively only represents a classical outcome interchange $+1 \leftrightarrow -1$ on both sides. Thus also the data matrix given by Eq.~(\ref{eq:exD3}) with marginals $-(1-\sqrt{3})$ is separable. However, taking the equal mixture leads to the ``depolarized'' data matrix $\tilde D_3 = \diag[1, D_2]$ which would verify entanglement according to Prop.~\ref{prop:main_result}.
\end{proof}

The following proposition demonstrates that the entanglement properties are not fully determined by the singular values of the observed data matrix. Moreover, for this special data structure it is interesting to observe that the extra knowledge of sharpness and orthogonality is irrelevant for the detection strength and mere information about the dimension of the measurements suffices to verify the same fraction of entanglement. Furthermore, since these observations satisfy all CHSH inequalities~\cite{clauser74a}, complete device independent detection is not possible.
\begin{proposition}(Diagonal data matrix)\label{prop:D3diag}
Observations of a diagonal data matrix
\begin{equation}
D_3 = \left[ \begin{array}{ccc} 1& &\\ & \lambda_1 & \\ & & \!\!\lambda_2 \end{array} \right],
\end{equation}
with $1\geq \lambda_{1/2} \geq 0$ verify entanglement under the assumption of qubit measurements if and only if $\lambda_1 + \lambda_2 >1$. Hence one verifies the same fraction as with sharp, orthogonal qubit measurements. In contrast, complete device independent entanglement verification fails.
\end{proposition}

\begin{proof}
The proof runs analogous to the qubit measurement scenario of Prop.~\ref{prop:main_result}. Note that the full correlation matrix $T_3$ is given by Eq.~(\ref{eq:general_T3}) and that one detects entanglement if and only if the singular values of $\bar T_2 = R_2(\alpha) \bar D_2 R_2(\beta)^T$ given by Eq.~(\ref{eq:general_T2bar}), satisfy  $\bar t_1 +\bar t_2 >1$. However in contrast to Prop.~\ref{prop:main_result} it is now possible to derive a tighter lower bound by exploiting the diagonal structure of
\begin{equation}
\bar D_2 = \left[ \begin{array}{cc} x_2 y_2 \lambda_1 & \\ & y_2 y_4 \lambda_2 \end{array} \right] = \left[ \begin{array}{cc} \bar \lambda_1 & \\ & \bar \lambda_2 \end{array} \right].
\end{equation}
The singular values of $\bar D_2$ are given by the diagonal entries, which fulfil $\bar \lambda_i \geq \lambda_i$, due to the constraints on the parameters $x_i,y_i$. In order to finish the proof we employ as usual certain inequalities for singular values of the matrices $\bar T_2$ and $D_2$:
\begin{align}
\bar t_1 \bar t_2 &\geq \lambda_1 \lambda_2, \label{eq:Prop331}\\
\bar t_1^2 + \bar t_2^2 & \geq \lambda_1^2+\lambda_2^2.\label{eq:Prop332}
\end{align}
These inequalities imply $\bar t_1 + \bar t_2 \geq \lambda_1 + \lambda_2$ and therefore prove the claim of the proposition.

In order to prove Ineq.~(\ref{eq:Prop331}) one applies the determinant multiplication rule together with the property $|\det[ R_2(\cdot)]| \geq 1$ that can be checked directly from the definition given by Eq.~(\ref{eq:R3}). The second ineq.~(\ref{eq:Prop332}) is verified by
\begin{align}
\bar t_1^2 &+ \bar t_2^2 = \tr( \bar T_2 \bar T_2^T) \\
\nonumber
=& \frac{1}{16}\left\{(\bar \lambda_1 \!+ \!\bar \lambda_2)^2 \!\! \left[ \csc(\alpha)^2\!\csc(\beta)^2 \!+\!\sec(\alpha)^2\!\sec(\beta)^2 \right] \right.\\
&\left.+(\bar \lambda_1 \! -\! \bar \lambda_2)^2 \!\!  \left[ \csc(\alpha)^2\!\sec(\beta)^2 \! + \! \sec(\alpha)^2 \! \csc(\beta)^2 \right] \right\} \\
\geq& \frac{1}{2}\left[ (\bar \lambda_1 + \bar \lambda_2)^2 + (\bar \lambda_1 - \bar \lambda_2)^2 \right] \geq \lambda_1^2 +\lambda_2^2,
\end{align}
where the first inequality is obtained by minimizing each term within the squared brackets separately.

Complete device independent entanglement verification fails because all possible Bell inequalities are satisfied, which is equivalent to a separable quantum representation in the bipartite case~\cite{werner89a,acin06a}.
\end{proof}

\section{Qutrits and beyond}\label{sec:IV}

As shown in the previous section solely having the knowledge that one is measuring a qubit is enough to detect entanglement even if the corresponding Bell inequalities, the set of inequivalent CHSH inequalities, are not violated. Nevertheless if all these Bell inequalities are satisfied then the observed data can be reproduced by appropriate measurements onto a higher dimensional separable state~\cite{werner89a,acin06a}; for the considered case this would be in dimensions $4 \otimes 4$. Thus the only other non-trivial case is the instance of qutrits. In the following we prove that even the qutrit assumption alone suffices to detect more than with the CHSH Bell inequalities.

Rather than defining different notions of sharpness or orthogonality for higher dimensional measurements or more settings, we focus on the completely uncharacterised case of $n$ dichotomic measurements on a $d$ dimensional system. Each dichotomic measurement is uniquely determined by the operator given by the difference of two POVM elements and is denoted as $A_i$ with $i=1,\dots,n$ in the following. The only defining inequality for all these operators $A_i$, besides that they are all acting on the same $d$-dimensional Hilbert space $\mathbbm{C}^d$, is the condition of Eq.~(\ref{eq:condpovm}) which ensures that they correspond to valid quantum measurements. Similar conditions are imposed for the measurements for Bob labelled as $B_i$. We employ once more the notion of a data matrix $D$, each entry defined as $[D]_{ij} = \mean{ A_i \otimes B_j}$ for $i,j=0,\dots,n$, with  $A_0=B_0=\mathbbm{1}$, such that it also contains the observed marginals. Obviously we also have $[D]_{00}=1$, which we always assume to be fulfilled if we speak about a data matrix. This is again employed to provide a more compact solution, which is stated in the following proposition. Besides from the above mentioned qutrit example in the CHSH case it has a few more consequences which are commented afterwards. The condition for qubit and qutrits are plotted in Fig.~\ref{fig:detectioin_regions}.

\begin{proposition}(Data matrix for $n$ dichotomic measurements on two qudits)\label{prop:detectMe}
If the data matrix $D$ corresponding to $n$ dichotomic measurements satisfies
\begin{equation}
\label{eq:detme}
|\det(D)| > \left( \frac{d}{n+1} \right)^{n+1}
\end{equation}
one verifies entanglement under the assumption of $d$-dimensional measurements. 

In case one has at least as much settings as dimensions, \ie, $n\geq d$, already the condition
\begin{equation}
\label{eq:con11}
|\det(D)| > \left( \frac{d-1}{n} \right)^{n}
\end{equation}
ensures entanglement. For the special case of two settings $n=2$ and qutrits $d=3$ the condition can be improved to
\begin{equation}
|\det(D)| > \frac{64}{81}.
\end{equation}
\end{proposition}

\begin{proof}
We follow a similar proof technique as in the previous section. The data matrix is transformed with appropriate corrections $G_a, G_b$ in order to form a kind of correlation matrix $C=G_a D G_b^T$ for which one employs a known entanglement criterion.

This correlation matrix $C$ is very similar to the previously employed matrix $T_3$. Here it is defined as $C[\rho]_{ij}=\tr(\rho K_i^{\rm A} \otimes K_j^{\rm B})$ where each local set $K_i$ consists of orthonormal (with respect to the Hilbert Schmidt inner product) observables, \ie, $\tr(K_i K_j) =\delta_{ij}$. Using the inclusion principle in a similar fashion as in the proof of Prop.~\ref{prop:ent_criterion} one finds that the state is entangled if the singular values of $C$, denoted as $c_i$, fulfil
\begin{equation}
\label{eq:CCN2}
\| C \|_1 = \sum_{i=0}^n c_i > 1.
\end{equation}

In the following we bound this trace norm by the determinant of the correlation matrix $|\det(C)|=\prod_i c_i$ and the extra knowledge that the largest singular value satisfies $c_0 \geq 1/d$ which is implied by the choice $K_0=\mathbbm{1}/\sqrt{d}$ and the inclusion principle. This provides the following estimate, 
\begin{align}
\sum_{i=0}^n c_i &\geq \min_{c_0 \geq 1/d} c_0 + \sum_{i=1}^n c_i \geq \min_{c_0 \geq 1/d} c_0 + n \left( \prod_{i=1}^n c_i \right)^{\frac{1}{n}} \!\!\!\! \\
\label{eq:opti_lowerbound_normC}
& \geq \min_{c_0 \geq 1/d} c_0 + \left( \frac{|\det(C)|}{c_0} \right)^{\frac{1}{n}} \\
\label{eq:lowerbound_normC}
& = \left\{ \!\! \begin{array}{cc} (n+1) |\det(C)|^{\frac{1}{n+1}} &  \text{if }\! |\det(C)|^{\frac{1}{n+1}} \geq \frac{1}{d} \\ \frac{1}{d} + n \left( d |\det(C)| \right)^{\frac{1}{n}} & \text{else} \end{array} \!\! \right.\! .
\end{align}
Here we employed the inequality of arithmetic and geometric means in the second step, while the optimization given by Eq.~(\ref{eq:opti_lowerbound_normC}) is performed using standard analysis. Note that the first solution in Eq.~(\ref{eq:lowerbound_normC}) is the unconstrained optimum that could have been inferred directly by applying the inequality of arithmetic and geometric mean to all terms. However, under the constraint on the largest singular value this solution is only reached if the determinant of the correlation matrix satisfies the stated extra condition. This distinction is necessary for the improved condition in case $n\geq d$.

The remaining strategy is to lower bound each solution of Eq.~(\ref{eq:lowerbound_normC}) by an expression involving the data matrix. Afterwards one investigates which conditions assure that this lower bound actually exceeds one such that, in spirit of Eq.~(\ref{eq:CCN2}), it would signal entanglement. The more stringent of these two cases will be the final entanglement criterion. Here we need the following inequality that relates the determinant of correlation and data matrix, 
\begin{align}
\label{eq:CDbound1}
|\det(C)|&=|\det(D)| \: |\det(G_a)| \: |\det(G_b^T)|\\
\label{eq:CDbound}
&\geq |\det(D)|d^{-(n+1)}
\end{align}
which follows from the bound $|\det(G)| \geq d^{-\frac{n+1}{2}}$ for each of the above mentioned transformation $G_a,G_b$, proven in App.~\ref{sec:boundforqutrit}. 

Let us start with the second solution and employ Eq.~(\ref{eq:CDbound}) which results in
\begin{equation}
\label{eq:cond22}
\frac{1}{d} + n \left( d |\det(C)| \right)^{\frac{1}{n}} \geq \frac{1}{d}\left( 1 + n |\det(D)|^{\frac{1}{n}} \right),
\end{equation}
which is larger than one if and only if the condition given by Eq.~(\ref{eq:con11}) holds.
For the first solution one obtains 
\begin{equation}
\label{eq:heeeellllpp}
(n+1) |\det(C)|^{\frac{1}{n+1}} \geq \frac{n+1}{d} \max\left[1, |\det(D)|^{\frac{1}{n+1}}\right].
\end{equation}
The first part in the maximum follows from the region constraint on $|\det(C)|$, while the second part is obtained using Eq.~(\ref{eq:CDbound}). The maximum appears because both bounds are valid. The right hand side of Eq.~(\ref{eq:heeeellllpp}) is larger than one if already one the terms does so. For the general case one chooses the second part of this maximum, which is larger than one if and only if the determinant of the data matrix satisfies Eq.~(\ref{eq:detme}). Because this condition is weaker than the previous condition from Eq.~(\ref{eq:cond22}) this is the entanglement criterion for the general case. For the special configuration of $n\geq d$ we employ the first part of the maximum since it always exceeds one. Hence only the condition from Eq.~(\ref{eq:cond22}) is the relevant for this case, which proves the part of the special case $n\geq d$.

The improved condition for the qutrit case follows from a sharper lower bound on the transformation $G_a,G_b$ given by $|\det(G)| \geq\sqrt{3}/8$ which is proven in App.~\ref{sec:boundforqutrit}.
\end{proof}

It is worth to stress that one can employ Prop.~\ref{prop:detectMe} not only with the determinant of the full data matrix $D$, but also for each sub-determinant. This describes the case that certain measurement settings are left out, which is useful when two or more settings coincide or are linearly dependent in which case the determinant of the whole data matrix vanishes.

Interestingly Prop.~\ref{prop:detectMe} detects bound entanglement. In what follows we provide an explicit example of a PPT bound entangled state that is detected via the criterion given by Prop.~\ref{prop:detectMe}, \ie, solely by having knowledge about the underlying dimension. The state
\begin{align}
\label{eq:benatti_state}
\rho_{\rm BFP}=\frac{1}{6}\big(& \Phi^+_{\rm AB} \Psi^-_{\rm A' B'} + \Psi^+_{\rm AB}\Psi^+_{\rm A'B'}+ \Psi^-_{\rm AB}\Phi^-_{\rm A'B'} \\
\nonumber
&+\Phi^-_{\rm AB}\Psi^+_{\rm A'B'}+\Phi^-_{\rm AB}\Psi^-_{\rm A'B'}+\Phi^-_{\rm AB}\Phi^-_{\rm A'B'} \big)
\end{align}
with $\Phi^+,\dots, \Psi^-$ denoting the projectors onto the standard two-qubit Bell states, has been shown to be $4\otimes 4$ bipartite PPT bound entangled~\cite{benatti04a} under the splitting $\rm AA'|BB'$. Assume that one performs ``good'' enough measurements described by all traceless operators $A_{kl} = \sigma_k^{\rm A} \otimes \sigma_l^{\rm A^\prime}$ built up by tensor products of the identity and the Pauli operators, and the same measurements for Bob. Mixed with white noise $\rho(p)=(1-p)\rho_{\rm BFP} + p \mathbbm{1}/16$, the criterion given by Eq.~(\ref{eq:con11}) becomes
\begin{equation}
|\det(D)| = \left[\frac{(1-p)}{3}\right]^{15} > \left( \frac{3}{15} \right)^{15},
\end{equation}
and thus verifies entanglement as long as $p<2/5=0.4$. This values seems to coincide with the point where entanglement disappears, \ie, for $p\geq 0.41$ the state is separable using the method of Ref.~\cite{barreiro10a}. This detection capability represents a clear advantage over Bell inequalities, where it is still unknown if measurements on a bipartite bound entangled state could at all violate a Bell inequality, though recent results seem to falsify this belief~\cite{vertesi11a}.

To conclude this section we consider the two-qubit Werner states $\rho_{\rm W}(p)=(1-p) \Psi^- + p\mathbbm{1}/4$ and three orthogonal standard measurements, \ie, $\sigma_x, \sigma_y, \sigma_z$ for both sides. Then the above stated conditions verify entanglement as long as the white noise parameter satisfies $p<2/3\approx 0.67$, which coincides with the point where entanglement vanishes. This represents another important improvement with respect to Bell inequalities, since first there is hardly any good Bell inequality known that detects entanglement if $p>1-1/\sqrt{2}\approx 0.29$ and second, above $p \geq 7/12\approx 0.58$ it is known that no measurement would violate a Bell inequality~\cite{barrett02a}.

\section{Conclusion \& Outlook}\label{sec:V}

We have investigated the task of entanglement detection for cases where only some partial information about the performed measurements is known or assumed. The considered scenarios included properties like sharpness, orthogonality or where only the dimension of the underlying measurements is fixed. Via this extra information one verifies more data as resulting only from entangled states than in the totally device independent setting while still keeping a good detection strength in comparison to the fully characterized case.

There are many further research lines connecting from here: A thorough investigation of higher dimensional states and more measurement settings is clearly interesting in order to clarify the power but also the limitations of this intermediate approach. For that one should be aware that the current methods only represent first techniques towards these directions. Here alternative tools might be necessary, even an efficient numerical approach would be of great help. Another approach would be the investigation of detection methods for the multipartite case in a similar intermediate setting. Since our criterion rests on an entanglement criterion based on the correlation matrix this extension might be possible utilizing recent detection methods for genuine multipartite entanglement using the correlation tensor~\cite{vicente11a,laskowski11a}. Regarding explicit tasks, since our results show that one verifies even entanglement if the underlying Bell inequality is not violated this means that the lower bound on the concurrence given in Ref.~\cite{liang11a} could be improved. This partially characterized scenario might also be useful in order to obtain steering equalities which are more robust against calibration errors, in similar spirit as in Ref.~\cite{smith11a}. Finally it is tempting to apply these result also to quantum key distribution operated in a similar intermediate setting as described here, which has already been started in Ref.~\cite{pawlowski11a}. For that in particular Prop.~\ref{prop:D3diag} is interesting because it describes exactly the kind of observations that one expects in an entanglement based BB84 protocol. Since it states that one verifies the same fraction of entanglement as with totally characterized measurements, this hints that a very strong ``semi-device independent'' key rate could be obtained if one just possesses the knowledge that one measures a qubit. This would allow a much larger freedom in finding appropriate squash models for quantum key distribution since the measurements don't have to be fixed anymore~\cite{beaudry08a}.

\section{Acknowledgement}

We thank J.~I.~Vicente, O.~G{\"u}hne, B.~Jungnitsch and N.~L\"utkenhaus for stimulating discussions, in particular J.~I.~Vicente for lots of help regarding the more general case. This work has been supported by the FWF (START prize Y376-N16) and the EU (Marie Curie CIG 293993/ENFOQI). O.~G. is grateful for support from the Industry Canada and NSERC Strategic Project Grant (SPG) FREQUENCY.

\appendix

\section{Proof of Prop. \ref{prop:ent_criterion}}\label{proof:ent_criterion}

In this appendix we provide the proof of the CCNR criterion adapted to our partial information setting.

\begin{proof}
It is only necessary to consider the case of partial information, since the CCNR criterion is typically formulated in terms of the full density operator. Following Ref.~\cite{guehne06a} the CCNR criterion can be expressed as follows:

Let $T_4$ denote the complete correlation matrix of two qubits, which results from $T_3$ by adding the remaining Pauli operator for each local side. Suppose that its corresponding ordered singular values are denoted as $t_0 \geq t_1 \geq t_2 \geq t_3 \geq 0$. Then the CCNR criterion states that for any separable state these singular values fulfil $\sum_{i=0}^3 t_i \leq 2$.

According to the inclusion principle, cf. Corollary 3.1.3 of Ref.~\cite{horn91a}, the ordered singular values are lower bounded by the singular values of any sub-matrix. Thus one obtains, $t_i \geq \lambda_i$ for $i=0,1,2$, such that one arrives at
\begin{equation}
\lambda_0 + \lambda_1 + \lambda_2 \leq \sum_{i=0}^3 t_i \leq 2.
\end{equation}
Whenever this condition is violated the state must necessarily be entangled.

In case of vanishing marginals with $\lambda_0=1$ the condition transforms into $\lambda_1 + \lambda_2 \leq 1$. Note that the parameters $\lambda_{1/2} \leq 1$ are also the singular values of the sub-matrix $T_2$. Using similar techniques as presented in Ref.~\cite{horodecki96a} it is possible to fit a rotated Bell-diagonal separable state to these data. This is achieved along the following lines: First, consider another correlation matrix $T[\rho]_{\alpha \beta}=\tr(\rho \sigma_\alpha \otimes \sigma_\beta)$ which is built up by the standard Pauli operators $\{ \sigma_x,\sigma_y,\sigma_z \}$ for each local side. Assume that the given sub-matrix $T_2$ is precisely the upper-two block of such a correlation matrix, \ie, $T=\diag[ T_2, 0]$ filled with additional zero entries. The singular value decomposition of this correlation matrix is given by $T=O_a \Lambda O_b$ with singular values $\Lambda=\diag[\lambda_1,\lambda_2,0] \geq 0$ and $O_a, O_b$ being special orthogonal matrices of similar block-diagonal form, \ie, $O_a=\diag[\bar O_a, \pm 1]$ and an analogous form of $O_b$. Here note that $\bar O_a,\bar O_b$ are the, not necessarily special orthogonal matrices form the singular value decomposition of $T_2=\bar O_a \diag[\lambda_1, \lambda_2] \bar O_b^T$.

Next let us discuss the special case of a diagonal correlation matrix, \ie, $T[\rho] =\diag[\lambda_1,\lambda_2,0]$: These data correspond to a Bell-diagonal state, abstractly expressed as $\rho_{\rm bs}=\sum_i p_i \ket{bs_i}\bra{bs_i}$ with standard Bell states $\ket{bs_i}$ and appropriate weights equal to $p_i=(1\pm\lambda_1\pm\lambda_2)/4$ having all four combinations. The above stated condition ensures that all probabilities are indeed non-negative and are upper bounded by $1/2$, which ensure separability in this case~\cite{horodecki96a}.

For the general case one employs the relation that any special orthogonal transformed correlation matrix corresponds to some special unitary transformation on the level of quantum states~\cite{horodecki96a},
\begin{equation}
O_a T[ \rho ] O_b^T = T[ U_a \otimes U_b \rho U_a^\dag \otimes U_b^\dag].
\end{equation}
Since local unitary transformation do not change the entanglement properties, the appropriately transformed state $U_a \otimes U_b \rho_{\rm bs} U^\dag_a \otimes U^\dag_b$ is the actual separable state for the general correlation matrix $T$. Finally, this rotation property is exploited once more to support the assumption that $T_2$ is the sub-matrix of the first two rows of $T$, since any two orthogonal vectors can be rotated such that it matches these axis. This finally proves the claim.
\end{proof}

\section{Details for two qubit case}\label{sec:tech_det}

\subsection{Relations for sharp nonorthogonal measurements, Eqns.~(\ref{eq:help1}),(\ref{eq:help2})}\label{sec:proof_help12}
Equation~(\ref{eq:help1}) is a direct consequence of the determinant multiplication rule, \ie, $\det(AB)=\det(A)\det(B)$, and that the absolute value of the determinant is equal to the product of its singular values. In order to derive the second inequality one employs the singular value identities
\begin{align}
\sum_{i=1}^k \sigma_i(AB) &\leq \sum_{i=1}^k \sigma_i(A)\sigma_i(B), \\
\prod_{i=1}^k \sigma_i(AB) &\leq \prod_{i=1}^k \sigma_i(A)\sigma_i(B),
\end{align}
where $\sigma_i(\cdot)$ denotes the decreasing ordered singular values, cf. Ref.~\citep{horn91a}. Applying these identities to Eq.~\ref{eq:D2-1} leads to
\begin{align}
\lambda_1 + \lambda_2 \leq & b_1 \sigma_1([R_2(\alpha)]^{-1}T_2) + b_2 \sigma_2([R_2(\alpha)]^{-1}T_2) \\
\nonumber
\leq & b_1 \sigma_1([R_2(\alpha)]^{-1}T_2) \\
& + b_2 [ a_1 t_1+ a_2 t_2 - \sigma_1(R_2(\alpha)^{-1}T_2) ] \\
\leq & (a_1 b_1) t_1 + (a_2 b_2) t_2 \\ \leq & \sqrt{\left( a^2_1 b^2_1 + a^2_2 b^2_2\right) \left(t_1^2+t_2^2 \right)},
\end{align}
where the Cauchy-Schwarz inequality is applied in the last step.

\subsection{Relations for non-sharp orthogonal measurements, Eqns.~(\ref{eq:cond1})-(\ref{eq:cond3})}\label{sec:proof_conds123}

The first condition given by Eq.~(\ref{eq:cond1}) follows from the inclusion principle \cite{horn91a} using the first entry as a sub-matrix.

The last inequality Eq.~(\ref{eq:cond3}) holds because of the determinant multiplication rule,
\begin{align}
t_0 t_1 t_2 &= |\det[S(\vec x)]| |\det(D_3)| |\det[S(\vec y)]|  \\ & = |\det( S_x D_2 S_y^T)| = (x_2 x_4) (\lambda_1 \lambda_2) (y_2 y_4)  \\ & \geq \lambda_1 \lambda_2,
\end{align}
and the bounds on the appearing parameters, \eg, $x_2 \geq 1 + |x_1| \geq 1$.

In order to prove Eq.~(\ref{eq:cond2}) we apply the inclusion principle to a particular chosen $2\times 2$ sub-matrix. For this argument the matrix $\bar D_2 = S_x D_2 S_y^T$ attains special importance. First, note that the singular values of $\bar D_2$, denoted as $\bar \lambda_1 \geq \bar \lambda_2$, satisfy $\bar \lambda_i \geq \lambda_i$ because the transformations satisfy $S_x, S_y - \mathbbm{1}\geq 0$. Next, employ the singular value decomposition $\bar D_2 = U \Sigma V^T$ with $\Sigma=\diag[\bar \lambda_1 ,\bar \lambda_2]$. Since the singular values of $T_3$ remain invariant under orthogonal transformations, we apply appropriate orthogonal matrices to diagonalize $\bar D_2$, which leads to
\begin{equation}
\left[ \begin{array}{cc} 1 &  \\  & U^T \end{array}\right] T_3 \left[ \begin{array}{cc} 1 &  \\  & V \end{array}\right] = \left[ \begin{array}{cc} 1 & \bar y^T  \\ \bar x^T & \bar x \bar y^T + \Sigma \end{array}\right],
\end{equation}
with $\bar x = U^T x, \bar y = V^T y$. Using the sub-matrix formed by the first two rows and columns one obtains
\begin{equation}
\left[ \begin{array}{cc} 1 &  \bar y_1 \\ \bar x_1 & \bar x_1 \bar y_1 + \bar \lambda_1 \end{array}\right],
\end{equation}
which has determinant $\bar \lambda_1 \geq \lambda_1$. Then the inclusion principle directly states Eq.~(\ref{eq:cond2}).

\section{Optimization problems}\label{sec:app}

In this appendix we prove two lemmas concerning optimization problems appearing in the proof of Prop.~\ref{prop:main_result}. They are mainly given because of completeness of the manuscript.

\begin{lemma}\label{lemma1}
Suppose $\lambda_1, \lambda_2 \geq 0$. Then the solution of
\begin{equation}
\min_{\alpha,\beta} \left[ \frac{(\lambda_1 + \lambda_2)^2}{a_1^2b_1^2 + a_2^2b_2^2} +  2 \frac{\lambda_1 \lambda_2}{a_1 a_2 b_1 b_2}\right]
\end{equation}
with $a_1=\sqrt{2\cos(\alpha)^2} \geq a_2=\sqrt{2\sin(\alpha)^2}$ and similar for $b_i$ with another angle $\beta$ is
\begin{equation}
\label{eq:opti1_sol}
\frac{1}{4}\left(\sqrt{\lambda_1}+\sqrt{\lambda_2}\right)^4.
\end{equation}
\end{lemma}

\begin{proof}
First note that the ordering of the singular values does not modify the solution if the optimization is performed over the full period of each angle since  wrong ordering only leads to larger function values. Using the parametrization $\alpha=\gamma + \delta, \beta=\gamma - \delta$ simplifies the function to
\begin{equation}
\label{eq:opti1}
\frac{(\lambda_1 +\lambda_2)^2}{2+\cos(4\gamma) + \cos(4\delta)} + \frac{4 \lambda_1 \lambda_2}{|\cos(4\gamma)-\cos(4\delta)|}.
\end{equation}
Hence it effectively only depends on $u=\cos(4\gamma)$ and $v=\cos(4\delta)$ which are both in the interval $[-1,1]$. Note that the boundary of this feasible set is characterized by either $u,v$ taking on the extreme values of this interval. In the following we prove that the minima lies at this boundary.

Using the linear variable transformation $x=u+v$, $y=u-v$ changes the function to
\begin{equation}
\frac{(\lambda_1 +\lambda_2)^2}{2+x} + \frac{4 \lambda_1 \lambda_2}{|y|}.
\end{equation}
Now, taking partial derivatives, one finds that this function is decreasing with respect to $x$ in the interval $(-2,2]$ and depending on the sign of $y$ also decreasing in the $+y$ or $-y$ direction [the exceptional cases $y=0$ or $x=-2$ can be excluded since they are no minima]. Since the variables $x,y$ are bounded this shows that the minima is attained at the boundary. Going back to the form given by Eq.~\ref{eq:opti1} means that either $\cos(4\gamma)=\pm 1$ or $\cos(4\delta)=\pm1$. Here it does not matter which cosine is put to its extreme values such that one only needs to consider one of them. Only one of these values $\pm1$ leads to the solution, for the other one can directly verify that its solution is larger than given by Eq.~\ref{eq:opti1_sol}. Concluding, only the following function needs to be optimized over the angle $\psi$,
\begin{equation}
\frac{(\lambda_1 +\lambda_2)^2}{3+\cos(\psi)} + \frac{4 \lambda_1 \lambda_2}{1-\cos(\psi)}.
\end{equation}
This can be performed directly by looking for the vanishing derivatives and using only its real solutions. This finally leads to the solution given by Eq.~\ref{eq:opti1_sol}.
\end{proof}

\begin{lemma}\label{lemma2}
Suppose $\lambda_0 \geq \lambda_1 \geq \lambda_2 \geq 0$. Then the solution of
\begin{eqnarray}
\label{eq:lemma2}
\min && \mu_0 + \mu_1 +\mu_2 \\
\nonumber
\st && \mu_0 \geq \lambda_0, \; \mu_0 \mu_1 \geq \lambda_0 \lambda_1, \\
\nonumber
&&  \mu_0 \mu_1 \mu_2 \geq \lambda_0 \lambda_1 \lambda_2, \\
\nonumber
&& \mu_0 \geq \mu_1 \geq \mu_2 \geq 0,
\end{eqnarray}
is $\lambda_0 + \lambda_1 + \lambda_2$.
\end{lemma}

\begin{proof}
In order to prove the lemma let us first state the solution of the following sub-problem,
\begin{equation}
\label{eq:sub}
\min_{x\geq x_{\rm min} \geq 0} x + \frac{\lambda}{x} = \left\{ \begin{array}{ll} 2\sqrt{\lambda} & \text{if } \sqrt{\lambda} > x_{\rm min} \\
x_{\rm min} + \frac{\lambda}{x_{\rm min}} & \text{else}\end{array} \right. ,
\end{equation}
with $\lambda > 0$ that follows using standard analysis. Note that $\sqrt{\lambda}$ is the argument of the unconstrained minima.

Now we turn to the intended problem given by Eq.~\ref{eq:lemma2}. At first consider the case that the parameter $\mu_0$ is fixed and that we bound the sum of the remaining two parameters from below, which provides
\begin{align}
\min_{\mu_1,\mu_2} \mu_1 + \mu_2 \geq \min_{\mu_1 \geq \bar \mu_1} \mu_1 + \left( \frac{\lambda_0 \lambda_1 \lambda_2}{\mu_0}\right) \frac{1}{\mu_1} \\
\label{eq:lower_bound}
\geq \left\{ \begin{array}{ll} 2 \sqrt{\frac{\lambda_0 \lambda_1 \lambda_2}{\mu_0}} & \text{if }\mu_0 \geq \bar \mu_0 \\
\frac{\lambda_0 \lambda_1}{\mu_0} +\lambda_2 &  \text{if } \lambda_0 \leq \mu_0 \leq  \bar \mu_0  \end{array} \right. ,
\end{align}
using the abbreviations $\bar \mu_1=\lambda_0 \lambda_1 / \mu_0$ and $\bar \mu_0 =\lambda_0 \lambda_1 /\lambda_2$. In the first inequality we employ the lower bound on $\mu_2$ from the problem formulation. The second inequality is an application of the sub-problem in which the conditions are re-expressed in terms of the parameter $\mu_0$. In the following we use these lower bounds and consider variations of $\mu_0$ within the corresponding valid region. For simplicity let us assume strict inequality $\lambda_0 > \lambda_1 > \lambda_2$; the statement with equality of some or all parameters follows then by continuity.

Let us start with the case $\lambda_0 \leq \mu_0 \leq \bar \mu_0$. Using the derived lower bound gives
\begin{align}
\min_{\substack{\mu_1, \mu_2 \\ \lambda_0 \leq \mu_0 \leq \bar \mu_0}}  \mu_0 &+ \mu_1 + \mu_2 \geq \min_{\mu_0\geq \lambda_0} \mu_0 + \frac{\lambda_0 \lambda_1}{ \mu_0 } +\lambda_2 \\
& \geq \lambda_0 + \lambda_1 + \lambda_2,
\end{align}
via another application of the given sub-problem. Note that the unconstrained minima is not achieved, \ie, $\sqrt{\lambda_0 \lambda_1} \not > \lambda_0$ because of the strict ordering.

Next consider the case $\mu_0 \geq \bar \mu_0$, for which we have to employ the other lower bound in Eq.~\ref{eq:lower_bound}. This leads to the
\begin{align}
\label{eq:sub1}
\min_{\substack{\mu_1, \mu_2 \\ \mu_0~\geq~\bar\mu_0}}  \mu_0 &+ \mu_1 + \mu_2 \geq \min_{\mu_0\geq \ \bar \mu_0} \mu_0 + 2 \sqrt{\frac{\lambda_0 \lambda_1 \lambda_2}{ \mu_0 }} \\
& \geq  \frac{\lambda_0 \lambda_1}{\lambda_2} + 2 \lambda_2 > \lambda_0 + \lambda_1 + \lambda_2.
\end{align}
The optimization problem appearing in Eq.~\ref{eq:sub1} is very similar to our sub-problem, in particular it is convex again for positive $\mu_0$. Its unconstrained minima is at the argument $\sqrt[3]{\lambda_0 \lambda_1 \lambda_2}$ which, however, is outside the allowed region, \ie, $\bar \mu_0 > \sqrt[3]{\lambda_0 \lambda_1 \lambda_2}$ due to the strict ordering of the $\lambda$'s. Thus the optima is attained at the boundary $\mu_0=\bar \mu_0$. The last inequality represents another consequence of the ordering property since it is equivalent to $\lambda_0 (\lambda_1-\lambda_2)  > \lambda_2 (\lambda_1-\lambda_2)$.
\end{proof}

\section{Transformation determinants}\label{sec:boundforqutrit}

In this appendix we provide a proof for the bounds on the determinant of the transformations $G$ used in the proof of Prop.~\ref{prop:detectMe}. It is a direct consequence of the Gram--Schmidt procedure.
\begin{lemma}
The linear transformation $G$ that maps the operators from the data matrix $\{ A_i \}$ all acting on $\mathbbm{C}^d$ with $K_0=\mathbbm{1}$ and  $-\mathbbm{1} \leq A_i \leq \mathbbm{1}$ for all $i=1,\dots,n$ into an orthonormal operator set $\{ K_i \}$, \ie, $\tr( K_i K_j) =\delta_{ij}$ fulfils
\begin{equation}
|\det(G)| \geq d^{-\frac{n+1}{2}}.
\end{equation}
For the case $n=2, d=3$ the bound can be improved to 
\begin{equation}
|\det(G)| \geq \frac{\sqrt{3}}{8}.
\end{equation}
\end{lemma}

\begin{proof}
The linear operation $G$ can be obtained for instance using the Gram-–Schmidt process. Without loss of generality this procedure can be decomposed into two operations $G=O N$. The first transformation $N$ should map each operator $A_i$ to its normalized form $\tilde  A_i=A_i/\sqrt{\tr(A_i^2)}$, such that the second transformation $O$ only needs to orthogonalize them. This second linear operation always stretches the ``vectors'' $\tilde  A_i$ such that the volume spanned by this set always increases, therefore $|\det(O)|\geq 1$. The first transformation $N$ is a diagonal matrix with entries given by $1/\sqrt{\tr(A_i^2)}$. Since each operator $A_i$ is bounded by the identity due to the restriction that it describes a valid measurement, one obtains $\tr{A_i^2} \leq \tr{\mathbbm{1}} \leq d$ or finally
\begin{equation}
|\det(G)| = |\det(O)|\: |\det(N)| \geq d^{-(n+1)/2}.
\end{equation}

For the special case of $n=2$ and $d=3$ we explicitly carry out the Gram--Schmidt process and minimize the determinant under the given constraints. We consider the case that $G$ maps the operators to following orthonormal set $\{ K_0=\mathbbm{1}/\sqrt{3}, K_1, K_2 \}$. This resulting operation $G$ is of triangular form, \ie,
\begin{equation}
G_a = \left[ \begin{array}{ccc} \frac{1}{\sqrt{3}} & & \\ \ast & N_1 & \\ \ast & \ast & N_2 \end{array} \right],
\end{equation}
which has a determinant $N_1 N_2 / \sqrt{3}$. In order to obtain a lower bound one needs to minimize $N_1 N_2$ or respectively maximize $1/(N_1 N_2)^2$, that results in
\begin{align}
\nonumber
&\left| \left[ \tr(A^2) - \frac{[\tr(A)]^2}{3}  \right] \left[ \tr(A^{\prime 2}) - \frac{[\tr(A^\prime)]^2}{3} \right] \right| \\ & \:\:\:\: - \left[ \tr(A A^\prime) - \frac{\tr(A)\tr(A^\prime)}{3} \right]^2 \\ & \leq \left| \left[ \tr(A^2) - \frac{[\tr(A)]^2}{3}  \right]  \left[ \tr(A^{\prime 2}) - \frac{[\tr(A^\prime)]^2}{3} \right] \right|  \\ & \leq \left( \frac{8}{3}\right)^2.
\end{align}
The last inequality originates from the bound
\begin{align}
\nonumber
&\tr(A^2) - \frac{[\tr(A)]^2}{3} \\
& \:\:\:= \frac{2}{3} \left( l_1^2 + l_2^2 + l_3^2 - l_1 l_2 - l_1 l_3 - l_2 l_3 \right) \leq  \frac{8}{3}
\end{align}
where $l_i$ are the eigenvalues of $A$ that satisfy a box constraint $|l_i| \leq 1$ due to the measurement condition of Eq.~\ref{eq:condpovm}. In this expression at most two of the last three terms can be positive but one must necessarily be negative. Suppose that this term is $-l_2l_3 < 0$, then $l_2^2 + l_3^2 -l_2 l_3 \leq 1$ holds given the stated box constraints.
\end{proof}

\bibliographystyle{apsrev}
%\bibliography{bibdiENT}

\end{document}